\documentclass[preprint,5p]{elsarticle}

\usepackage[toc,page]{appendix}
\usepackage{amssymb}
\usepackage{amsmath}
\usepackage{hyperref}
\usepackage{xcolor}

\usepackage{placeins} 
\usepackage{multirow} 
\usepackage{subcaption}

\newcounter{bla}

\journal{Computer Physics Communications}

\newcommand{\pdfflow}{\texttt{PDFFlow}~}
\newcommand{\vegasflow}{\texttt{VegasFlow}~}
\newcommand{\vegasflowpdfflow}{\texttt{VegasFlow-PDFFlow}~}

\usepackage{tikz}
\usetikzlibrary{decorations.pathmorphing,decorations.markings}
\usetikzlibrary{decorations.pathreplacing}
\usetikzlibrary{positioning, calc, shapes}

\usetikzlibrary{shapes.geometric, arrows, calc}
\tikzstyle{tool} = [rectangle, rounded corners, minimum width=1cm, text
					centered, draw=black, fill=red!30]
\tikzstyle{func} = [rectangle, dashed, minimum width=1cm, minimum height=0.6cm,
					text centered, draw=black, fill=green!30]
\tikzstyle{class} = [rectangle, rounded corners, minimum height=0.6cm, minimum
					 width=1cm,text centered, draw=black, fill=blue!30]
\tikzstyle{block} = [rectangle, dashed, minimum width=1cm, minimum height=0.6cm,
					 text centered, draw=black, fill=white]

\tikzstyle{arrow} = [thick,->,>=stealth]
\tikzstyle{line} = [thick,-,>=stealth]

\begin{document}

\begin{frontmatter}



\title{\texttt{PDFFlow}: parton distribution functions on GPU}


\author[label1]{Stefano Carrazza\corref{author}}
\author[label1]{Juan M.~Cruz-Martinez}
\author[label1,label2]{Marco Rossi}

\cortext[author] {Corresponding author.\\\textit{E-mail address:} stefano.carrazza@unimi.it\\\textit{Preprint number:} TIF-UNIMI-2020-19}
\address[label1]{TIF Lab, Dipartimento di Fisica,\\ Universit\`a degli Studi di Milano and
INFN Sezione di Milano,\\ Via Celoria 16, 20133, Milano, Italy}

\address[label2]{CERN openlab, Geneva 23, CH-1211, Switzerland}

\begin{abstract}
    We present \texttt{PDFFlow}, a new software for fast evaluation of parton
    distribution functions (PDFs) designed for platforms with hardware
    accelerators. PDFs are essential for the calculation of particle physics
    observables through Monte Carlo simulation techniques. The evaluation of a
    generic set of PDFs for quarks and gluon at a given momentum fraction and
    energy scale requires the implementation of interpolation algorithms as
    introduced for the first time by the LHAPDF project. \pdfflow
    extends and implements these interpolation algorithms using Google's
    TensorFlow library providing the capabilities to perform PDF evaluations
    taking fully advantage of multi-threading CPU and GPU setups. We benchmark
    the performance of this library on multiple scenarios relevant for the
    particle physics community.
\end{abstract}


\end{frontmatter}

\noindent
{\bf PROGRAM SUMMARY}
\\

\begin{small}
\noindent
{\em Program Title:} {\tt PDFFlow} \\
\\
{\em Program URL:} \url{https://github.com/N3PDF/pdfflow}\\
\\
{\em Licensing provisions:} GPLv3 \\
\\
{\em Programming language:} Python, C \\
\\
{\em Nature of the problem:} The evaluation of a generic set of parton
distribution functions requires the implementation of interpolation algorithms.
Currently, there are no public available implementations with hardware
acceleration support.\\
\\
{\em Solution method:} Implementation of interpolation algorithms for the
evaluation of parton distribution functions and the strong coupling $\alpha_s$
using the dataflow graph infrastructure provided by the TensorFlow framework,
taking advantage of multi-threading CPU and GPU setups. \\

\end{small}


\section{Introduction and motivation}
\label{sec:introduction}

Parton Distribution Functions (PDFs) are at the center of High Energy Physics
(HEP) phenomenology by providing a description of the parton content of the
proton which is an essential ingredient for the computation of physical
observables such as cross section and differential distributions. PDFs are
provided by fitting collaborations, each following different methodologies and
assumptions~\cite{Ball:2017nwa, Hou:2019jgw,Harland-Lang:2014zoa}. For its use
by the community a common format was agreed and adopted~\cite{Whalley:2005nh}, the LHAPDF
library~\cite{Buckley:2014ana} provides access to the PDFs as functions of the
momentum fraction of the parton and the energy scale. It provides interfaces
for Fortran, C/C++ and Python and has become a fundamental tool for most LHC-era
theoretical calculations.

The LHAPDF library has been developed with a focus on CPU computations,
originally single-threaded and now multi-threaded. The ever decreasing cost of
hardware accelerators such as GPUs, TPUs or FPGAs has given rise to new of
frameworks, such as \vegasflow~\cite{Carrazza:2020rdn,juan_cruz_martinez_2020_3691927}, which target said
devices in order to decrease the heavy computational cost of HEP phenomenology.
This has created the need for a hardware-agnostic PDF evaluation framework.
Moreover, the recent interest in Machine Learning techniques, which is highly
coupled with the use of hardware accelerators, could also benefit from a
hardware agnostic PDF evaluation framework.

The development of GPU-based phenomenology software has been regarded with expectation but
scepticism~\cite{Buckley:2019wov, Valassi:2020ueh} as most of the work has only
been shown in very particular simple cases such as Leading Order or inclusive
calculations. This should not be understood, however, as an actual limitation of the
available hardware but of the human resources available. A full new
calculation can take years to be completed, including a revamping of the
hardware and software used could add several years to it. In these
situations a choice is made, commonly favouring new physical results over
technical improvements~\cite{Valassi:2020ueh}. As a result, many of the most
precise and advanced calculations in the field are performed with dated software
which, despite being often nominally optimized for efficiency,
it is not designed to make full use of the current hardware.

In this paper we continue past work to provide the community with tools to
interface state of the art theoretical progress with hardware accelerators or
Machine Learning libraries. An important ingredient of higher order theoretical
predictions is the estimation of parton densities. They are relevant due to the
calculation of physical observables through the convolution of the matrix
element and the PDFs. As has been shown in the past, GPU-accelerated PDF
determination~\cite{Carrazza:2019agm} can considerably increase the efficiency
and performance of these computations.

We thus present the \pdfflow library~\cite{juan_cruz_martinez_2020_3964191}, where
the main contribution  is a  novel  implementation of the PDF and $\alpha_s$
interpolation algorithm used in LHAPDF able to run both in CPUs and GPUs,
enabling further acceleration of Monte Carlo simulation. The library is written
using the TensorFlow~\cite{tensorflow2015:whitepaper} framework hence the chosen
name. Our main goal with this publication is to bring the advantages of TensorFlow
(frictionless multi-hardware integration, analysis of the code into the XLA accelerated
linear algebra library) to
parton collision simulations with little to no effort made by the user and
developer.

This paper is structured as follows, in section~\ref{sec:techimp} we describe
the technical implementation of our library. In section~\ref{sec:bench} we
benchmark \pdfflow against LHAPDF from the point of view of precision and
performance. We show how indeed the usage of hardware accelerators can greatly
reduce computational time for many applications. In section~\ref{sec:examples} we
show some implementation examples that can be useful for the broad HEP
community. We present Leading Order and Next-to-Leading Order examples
benchmarked against CPU-only code showing how GPUs are already at a stage in
which they are competitive and can surpass professional grade CPU hardware, even
with little to no GPU specific optimization. Then we compare the performance of
\pdfflow when evaluating multi-PDF members in a $x$-grid of points similarly to
the \texttt{FastKernel} procedure presented in~\cite{Bertone:2016lga} which has
already being shown to be much faster when running on
accelerators~\cite{Carrazza:2019mzf}. Finally, in section~\ref{sec:outlook} we
present our conclusion and future development direction.

\section{Technical Implementation}
\label{sec:techimp}

\subsection{\pdfflow design}

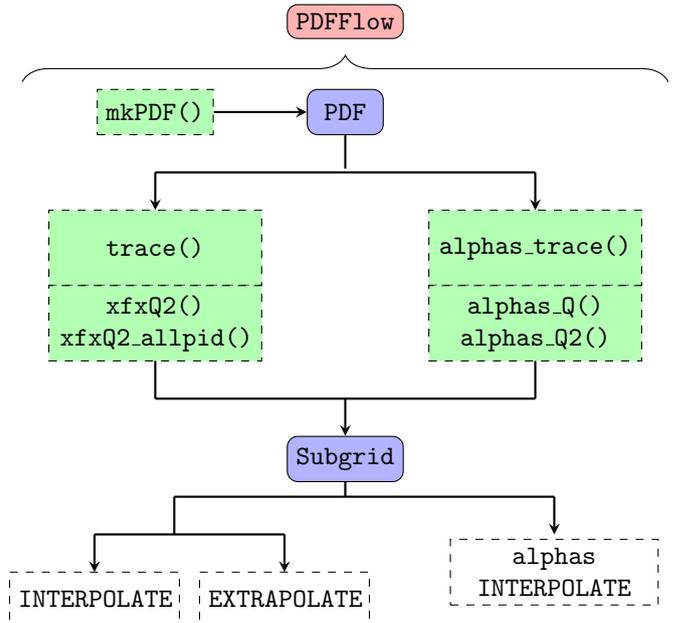
\begin{figure}
	\begin{tikzpicture}[node distance = 1cm]
	\centering
	\node[tool] (pdfflow) at (4.25,0) {\pdfflow};

	\node[class] (PDF) at ($(pdfflow)-(0,1.2)$) {\texttt{PDF}};
	\node[func] (mkPDF) at ($(PDF)-(2.5,0)$) {\texttt{mkPDF()}};

	\node[func, minimum height=1cm, minimum width=2.8cm] (trace)
	at ($(pdfflow)-(2.5,3)$){\texttt{trace()}};
	\node[func, text width=2.5cm, minimum height=1cm, minimum width=2.8cm]
	(xfxq2) at ($(trace)+(0,-1)$){\texttt{xfxQ2()\\xfxQ2\_allpid()}};

	\node[func, minimum height=1cm, minimum width=2.8cm] (atrace)
	at ($(pdfflow)+(2.5,-3)$) {\texttt{alphas\_trace()}};
	\node[func, text width=2.5cm, minimum height=1cm, minimum width=2.8cm]
	(axfxq2) at ($(atrace)+(0,-1)$) {\texttt{alphas\_Q()\\alphas\_Q2()}};

	\node[class] (subgrid) at ($(xfxq2)+(2.5,-1.8)$) {\texttt{Subgrid}};

	\node[block, text width=2.5cm, ] (alphasint) at
	($(subgrid)+ (2.75,-1.5)$) {\texttt{alphas\\INTERPOLATE}};

	\node[block, text width=2cm, minimum height=0.7cm] (interp)
	at ($(subgrid)+(-3.3,-1.85)$) {\texttt{INTERPOLATE}};
	\node[block, text width=2cm, minimum height=0.7cm] (extrap)
	at ($(interp)+(2.5,0)$) {\texttt{EXTRAPOLATE}};

	\draw [decorate,decoration={brace,amplitude=10pt}]
	($(pdfflow)-(4.25,0.8)$) -- ++(right:8.5);

	\draw[arrow] (mkPDF) -- (PDF);

	\draw[line] (PDF) -- ++(down:0.75);
	\draw[line] ($(trace)+(0,1)$) -- ($(atrace)+(0,1)$);
	\draw[arrow] ($(trace)+(0,1)$) -- ++(down:0.5);
	\draw[arrow] ($(atrace)+(0,1)$) -- ++(down:0.5);

	\draw[line] ($(xfxq2)-(0,0.5)$) -- ++(down:0.5);
	\draw[line] ($(axfxq2)-(0,0.5)$) -- ++(down:0.5);
	\draw[line] ($(xfxq2)-(0,1)$) --($(axfxq2)-(0,1)$);
	\draw[arrow] ($(xfxq2)+(2.5,-1)$) -- (subgrid);

	\draw[line] (subgrid) -- ++(down:0.5);
	\draw[line] ($(alphasint)+(0,1)$) -- ++(left:5);
	\draw[arrow] ($(alphasint)+(0,1)$) -- ++(down:0.5);

	\draw[line] ($(alphasint)+(-5,1)$) -- ++(down:0.5);
	\draw[line] ($(interp)+(0,0.85)$) -- ($(extrap)+(0,0.85)$);
	\draw[arrow] ($(interp)+(0,0.85)$) -- ++(down:0.5);
	\draw[arrow] ($(extrap)+(0,0.85)$) -- ++(down:0.5);
	\end{tikzpicture}
	\caption{\pdfflow flowchart. Blocks are color-coded as following: red for
	the tool, violet for classes, green for functions and class methods,
	white for interpolation algorithms.}
	\label{fig:flowchart}
\end{figure}

Figure~\ref{fig:flowchart} depicts the \pdfflow design. The \texttt{mkPDF()}
function instantiates the desired PDF representation, given by the \texttt{PDF}
class. A \texttt{PDF} object stores all the quantities and algorithms needed
for the interpolation both the PDF itself and the strong running coupling
$\alpha_s$. Notable member methods in the class are the trace methods
(\texttt{trace} and \texttt{alphas\_trace}) and interpolating methods (two for
the PDF and two for $\alpha_s$), contained in the green dashed boxes in the
figure.

The key concept in \pdfflow is the implementation of the interpolating methods:
these functions are decorated by the \texttt{tf.function} method, that triggers
the \texttt{tf.Graph} computation and ensures hardware acceleration.
\texttt{Tensor-\\Flow} converts Python code into \texttt{tf.Tensor} and
\texttt{tf.Ope-\\ration} primitives, when it executes new code. This conversion
procedure introduces an overhead in the running time and in order to optimize
it as much as possible, \pdfflow provides the user with trace methods that allow
ahead of time compilation of all the functions declared within the tool.

The interpolating methods include a call to a \texttt{Subgrid} class object.
\texttt{Subgrid} stores PDF grid data and includes a switch to enable
interpolation on $\alpha_s$ grid knots. This class gives access to algorithms
that implement the actual computation of \pdfflow outputs, represented by white
boxes in the flowchart and briefly described in sections~\ref{subsec:pdf eval}
and~\ref{subsec:alphas eval}.
The resemblance to the structure of LHAPDF is, of course, not a coincidence
as it has been the standard PDF estimation library for many years.

\subsection{PDF evaluation procedure}
\label{subsec:pdf eval}

The interpolation procedure implemented in \pdfflow follows the prescription
originally implemented in LHAPDF6, namely a log-bicubic PDF interpolation in
terms of $x$ and $Q^2$, respectively the parton momentum fraction and the
reference scale.
The PDF data files stored in the LHAPDF directories are directly loaded into
\texttt{tf.Tensor} objects so it is not necessary to install new sets or formats.
The interpolation algorithms computes independently query
points belonging to different sub-grids of the PDF set. Special care is taken
about regions in the $(x,Q^2)$ plane close to quark mass thresholds and grids
$x$ edges, where the minimum number of knots required for bicubic interpolation
is not available.

\subsection{$\alpha_s$ evaluation procedure}
\label{subsec:alphas eval}

Similarly to the PDF interpolation procedure, the evaluation of the running
of the strong coupling, $\alpha_s(Q)$ is performed using a log-cubic interpolation
with constant extrapolation from the $(\alpha_s(Q),Q)$ nodes stored in the PDF
metadata file. The implementation includes the improved treatment of the
sub-grids mechanism and takes into account the impact of
flavour thresholds on $\alpha_s(Q)$ evolution.

\section{Benchmarks}
\label{sec:bench}

\begin{table*}
    \centering
    \begin{tabular}{c|l|c|c|c|c}
        Device & CPU model & CPU cores & CPU RAM & GPU(s) model & GPU memory \\ \hline
        C & Intel i7-6700K & 4 @ 4-4.2GHz & 16GB @ 3000MHz & Nvidia RTX2080 & 8GB\\
        \hline
        P0 & AMD 2990WX & 32 @ 3-4.2GHz & 128GB @ 3000MHz & - & - \\
        \hline
        P1 & Intel i9-9980XE & 18 @ 3-4.4GHz & 128GB @ 2666MHz& Nvidia TITAN V & 12GB\\
         &  &  &  & Nvidia RTX2080TI & 12GB\\
         \hline
        P2 & Intel Xeon Gold 6126 & 6 @ 2.6-3.7GHz & 20GB @ 2133MHz & Nvidia V100 (2x) & 32GB\\
        \hline
    \end{tabular}
    \caption{Description of the systems in which the different codes have been run.}
    \label{table:hardware}
\end{table*}

In this section we present benchmark results between \pdfflow v1.0 and {\tt LHAPDF
v6.3.0} libraries for the interpolation accuracy and performance.

\subsection{System configuration}

In table~\ref{table:hardware} we present a description of the hardware used in
this work, in particular for the performance examples in
section~\ref{sec:performance} and examples in section~\ref{sec:examples}.
The consumer grade hardware (C) consists in a standard desktop computer with
gaming level specifics. Different research groups have access to professional
grade hardware which is better suited for the kind of computation described in
this paper. In particular this corresponds to many-threaded CPUs and GPUs with
enough memory to hold the necessary kernels for very complicated computations.

For the CPU-based calculation we use the P0 system with a medium level processor
in terms of clock speed, while for the GPU-based calculations we use two
different machines: P1 with a very powerful processor, which greatly reduces the
latency of the calculation for CPU-based operation such as the accumulation of
the final results, and P2, a less powerful CPU and a more limited RAM size which
can add an important overhead on the communications between the CPU and the GPU.
In exchange the V100 GPUs have greater memory size which reduces the frequency
of the communications between the main memory and the device.

\subsection{Accuracy}
\label{sec:accuracy}

In order to measure and compare the PDF interpolation accuracy between \pdfflow
and LHAPDF, we define a relative difference:
\begin{equation}
    r_{i}(x,Q) = \frac{| x f_i^{\rm \pdfflow}(x,Q) - x f_i^{\tt LHAPDF}(x,Q)|}{| x f_i^{\tt LHAPDF}(x,Q)| + \epsilon},
    \label{eq:diff}
\end{equation}
where $x f_i (x, Q)$ is the numeric value of a PDF flavour $i$ evaluated at a
given momentum fraction $x$ and energy $Q$, and $\epsilon=10^{-16}$.

\begin{figure}[h]
    \centering
    \includegraphics[width=0.48\textwidth]{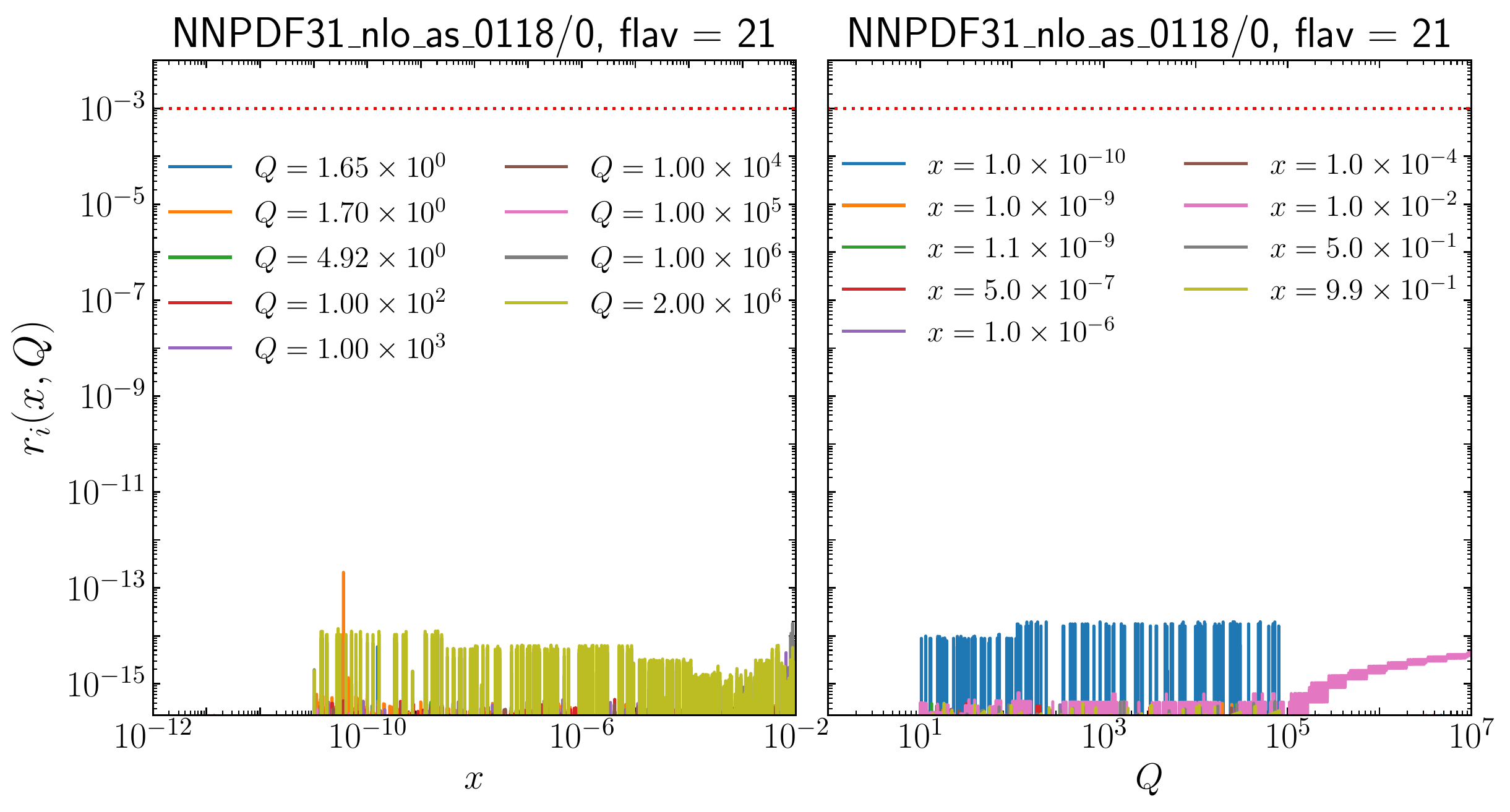}
    \includegraphics[width=0.48\textwidth]{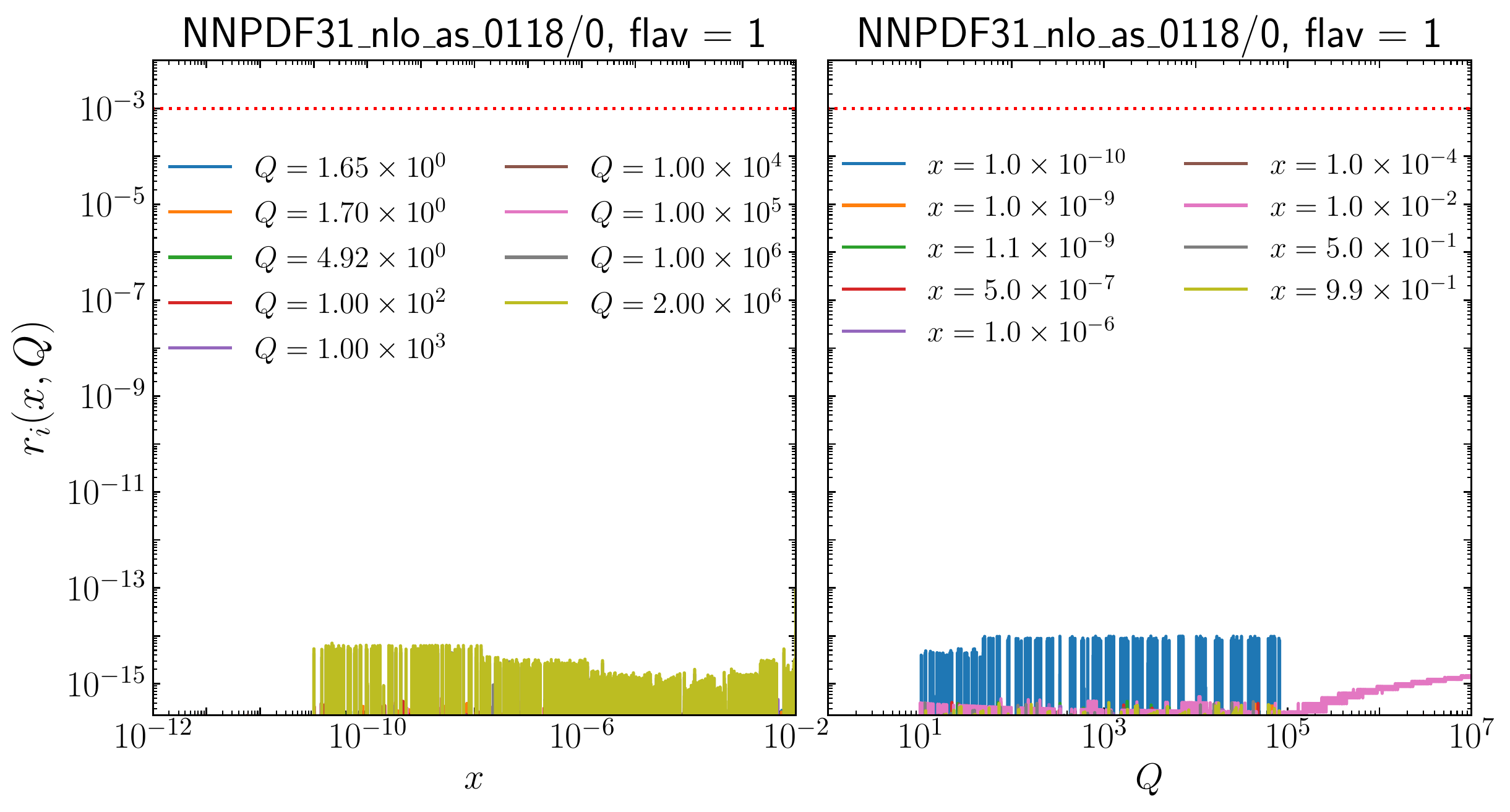}
    \caption{\pdfflow vs LHAPDF relative difference for the NNPDF3.1 NLO central
        value gluon PDF (first row) and $d$ quark PDF (second row). The first
        column refers to differences in a grid of $x$ points for fixed $Q$
        values, while the second column shows differences in a grid of $Q$
        values for fixed $x$.}
    \label{fig:pdfaccuracy_nnpdf}
\end{figure}

In figure~\ref{fig:pdfaccuracy_nnpdf} we show the relative difference in
eq.~\ref{eq:diff} for the NNPDF3.1 NLO set~\cite{Ball:2017nwa}. The first row
refers to the central member of the gluon PDF while the second row to the $d$ quark PDF.
The left column shows the difference for fixed $Q$ values in a grid of x points,
while the column on the right has fixed $x$ evaluated in a grid of $Q$ points.
For all curves we see that the relative difference is much below the nominal
$10^{-3}$ threshold that was used to compare LHAPDF5 to LHAPDF6
output~\cite{Buckley:2014ana}. Similar accuracy results are obtained for other
PDF sets, in particular in~\ref{app:pdfacuracy} we show an all-flavour
comparison for the MMHT2014 NLO PDF~\cite{Harland-Lang:2014zoa}.

\begin{figure}[t]
    \centering
    \includegraphics[width=0.48\textwidth]{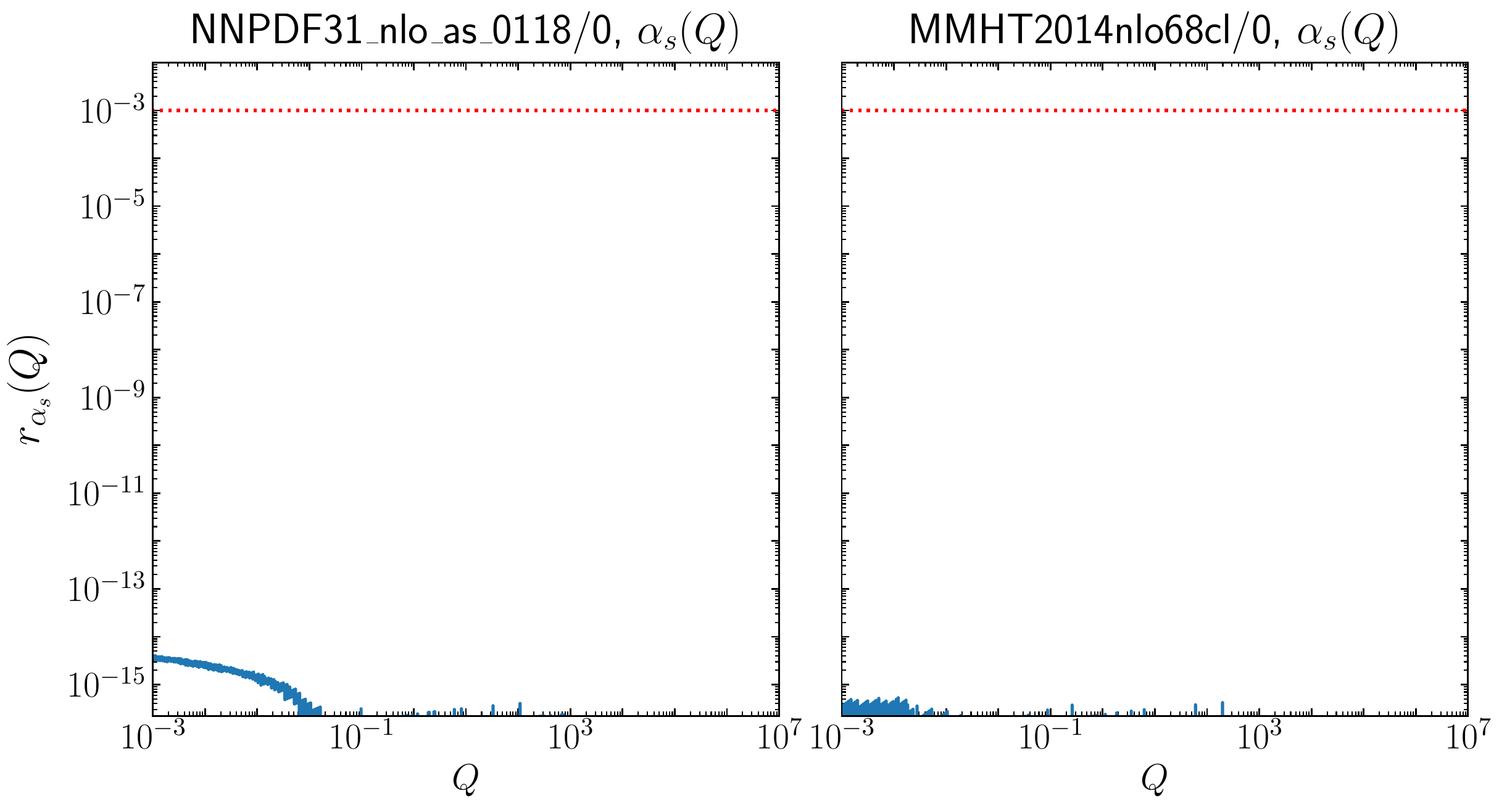}
    \caption{\pdfflow vs LHAPDF $\alpha_s$ relative difference for NNPDF3.1 NLO (left) and MMHT14 NLO (right),
    perfect agreement is found between the two codes.}
    \label{fig:asaccuracy}
\end{figure}

Concerning the $\alpha_s$ computation, in figure~\ref{fig:asaccuracy} we compute
the relative difference $r_{\alpha_s}(Q)$, using eq.~\ref{eq:diff}, by replacing
the $xf_i(x,Q)$ terms with $\alpha_s(Q)$, the strong coupling, as a
function of $Q$ points. We evaluate the relative difference in a grid of $Q$. We
observe that also in this case the relative differences are tiny and close to
the numerical precision of a 64-bits floating-point representation.
Note that the current \pdfflow implementation only supports $\alpha_s(Q)$
interpolated values as described in section~\ref{sec:techimp}.

\subsection{Performance}
\label{sec:performance}

\begin{figure}[t]
    \centering
    \includegraphics[width=0.4\textwidth]{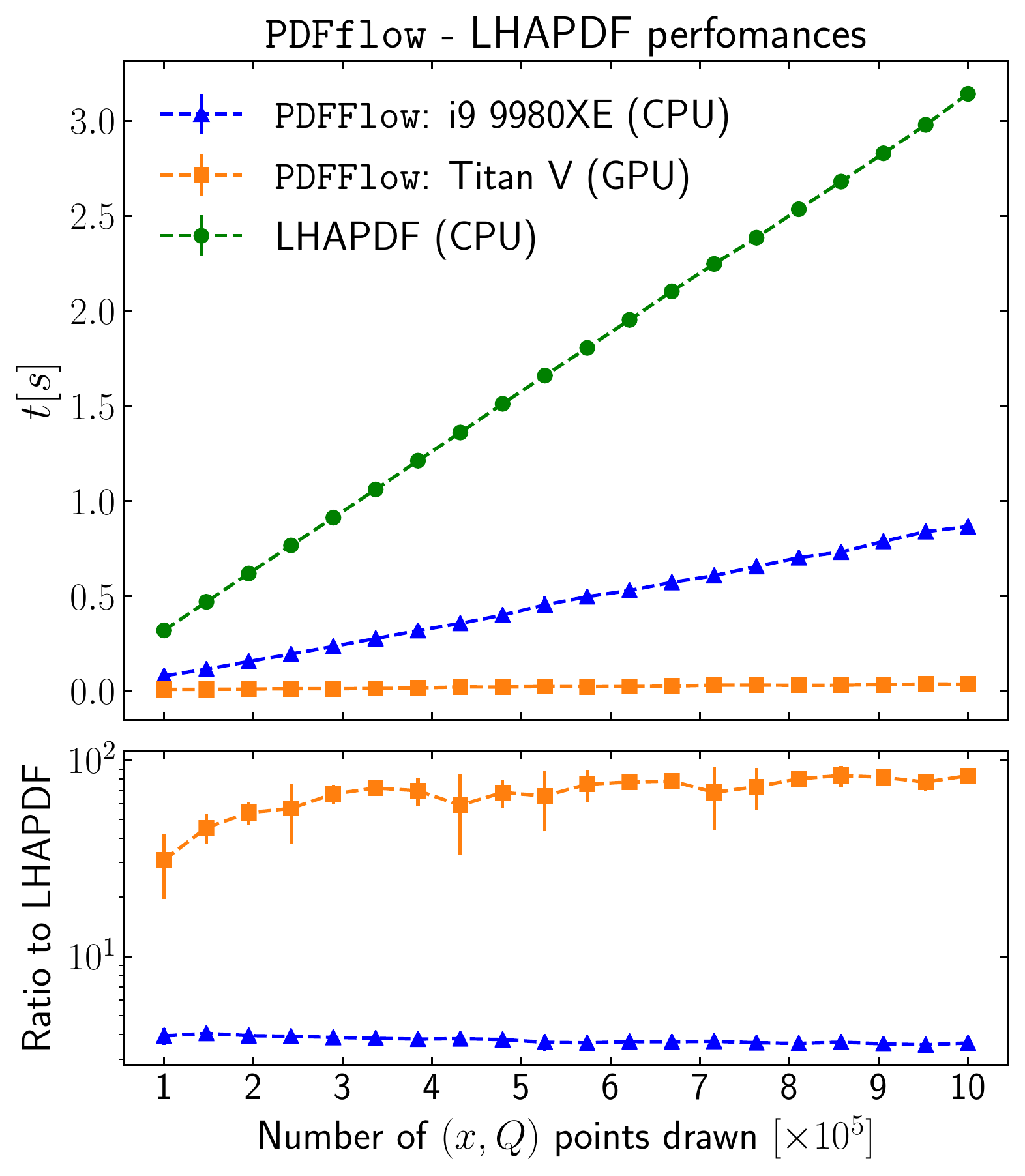}
    \caption{Top row: CPU and GPU performance on P1. Bottom row: performance ratio with LHAPDF on P1.}
    \label{fig:cpuperformance}
\end{figure}

In terms of performance speed, in figure~\ref{fig:cpuperformance} we compare
the evaluation time of \pdfflow on CPU and GPU to LHAPDF. The plot shows the
required time for \pdfflow and LHAPDF to perform the evaluation of all PDF
flavours for an increasingly large number of points distributed logarithmically in
a two-dimensional grid of $(x,Q)$ points, defined within the boundaries of the
original PDF grid.

We observe a great performance improvement when running the PDF query using
\texttt{PDFFlow}'s default configuration on CPU. Such improvement is due to the
built-in multi-threading CPU support. Concerning GPU results, the performance
improvement is massive and opens the possibility to construct new models and
applications with parallel evaluations.

\section{Examples}
\label{sec:examples}

In this section we first show examples of particle physics simulation
configurations where we integrate the \pdfflow library in \vegasflow~\cite{Carrazza:2020rdn,juan_cruz_martinez_2020_3691927},
a Monte Carlo integration framework for hardware accelerators. Then we compare the
performance of \pdfflow to {\tt LHAPDF} in the context of a multi-PDF member
evaluation.

\subsection{Single $t$-quark production at leading order}

\begin{figure}
    \centering
    \includegraphics[width=0.48\textwidth]{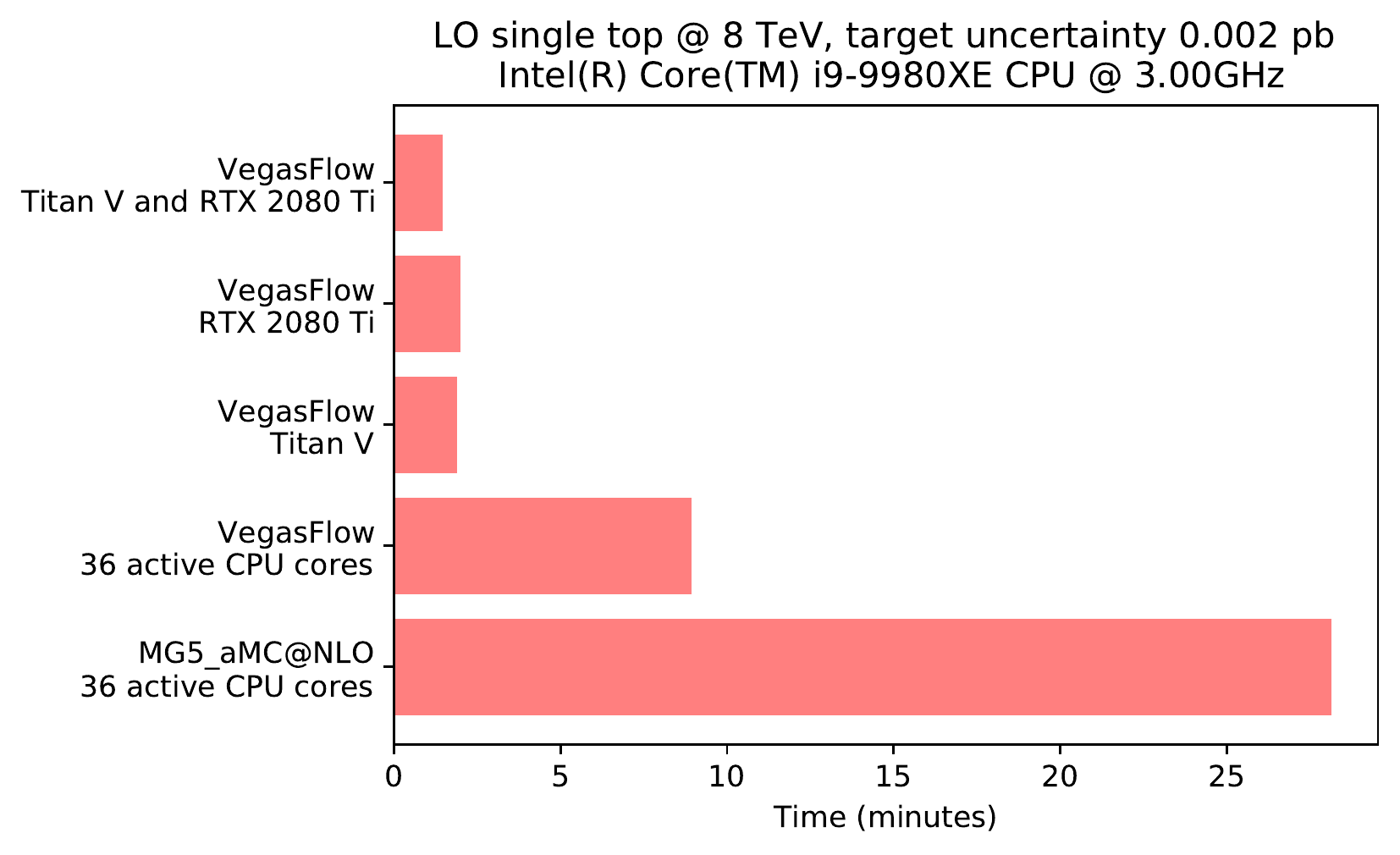}
    \caption{Comparison of a Leading Order calculation ran in both
    \vegasflowpdfflow and MG5\_aMC@NLO~\cite{Alwall:2014hca}. The CPU-only
    version of \vegasflowpdfflow is able to improve the performance obtained by
    MG5\_aMC@NLO for the same level of target precision. The usage of the GPU
    devices further improves the performance.}
    \label{fig:singletop}
\end{figure}

As a first proof of concept we have implemented a Leading Order (LO) integration
of single $t$-quark production (t-channel) at the partonic level using the
\vegasflow and \pdfflow libraries. This approach is optimal given that the
evaluation of Monte Carlo events is performed in parallel, thus the PDF
evaluation is parallelizable.
We compare our calculation with predictions from
MG5\_aMC@NLO~\cite{Alwall:2014hca} at LO, using the same physical parameters
such as the $t$-quark mass, $m_t=173.2$ GeV  and centre of mass energy
$\sqrt{s}= 8$ TeV.

In figure~\ref{fig:singletop} we compare the total execution time for
\vegasflowpdfflow on single GPU, multi-GPU and multi-threading CPU
configurations to the equivalent fixed LO order provided by MG5\_aMC@NLO 3.0.2.
In both cases we use the central replica from the NNPDF3.1 NLO set. The stopping
criteria for the total number of events relies on a target precision of
$2\cdot10^{-3}$ pb (relative error of 0.004\%). We observe great improvement in terms of execution time for
the \vegasflowpdfflow approach.

\subsection{Higgs production on VBF at Next-to-Leading Order}

In order to show the full potential of GPU computing, in the next paragraphs we
present a Next-to-Leading Order (NLO) implementation of Vector Boson Fusion
(VBF) Higgs production.

A full implementation of a parton-level Monte Carlo simulator such as
NNLOJET~\cite{Gehrmann:2018szu} or MCFM~\cite{Campbell:2019dru} is beyond the scope of this work,
which is to provide a proof of concept for a NLO computation.
We implement a simplified version of the existing Fortran 95 implementation
of the process found in NNLOJET~\cite{Cruz-Martinez:2018rod}, which uses LHAPDF
as the PDF provider.
We limit ourselves to the
quark initiated W-boson mediated process, with a gluon radiated from any of the
quarks at NLO.
The integration of this process requires the implementation of a non-trivial
phase space, including a divergence that needs to be subtracted at NLO.

The divergent nature of the real radiation integrand requires the implementation of phase space cuts as well as
a subtraction scheme. We choose antenna subtraction to be comparable with the Fortran implementation.
This corresponds to the implementation of the two-quarks one-gluon antennae~\cite{GehrmannDeRidder:2005cm}
and the appropriate phase space mappings.

The NNLOJET code is heavily optimized for CPU and CPU-cluster usage so it provides a good benchmarking
ground for our python-TensorFlow implementation which is to be run on a GPU.
We do not change the phase space algorithms or subtraction strategies and so the implementation could be suboptimal
for GPU, however, the fact that we can achieve a level of performance which is competitive with NNLOJET lead us
to believe that a fully optimized implementation of NLO (and NNLO) computations on GPUs can lead to very important
performance gains.

\begin{figure}
    \centering
    \includegraphics[width=0.48\textwidth]{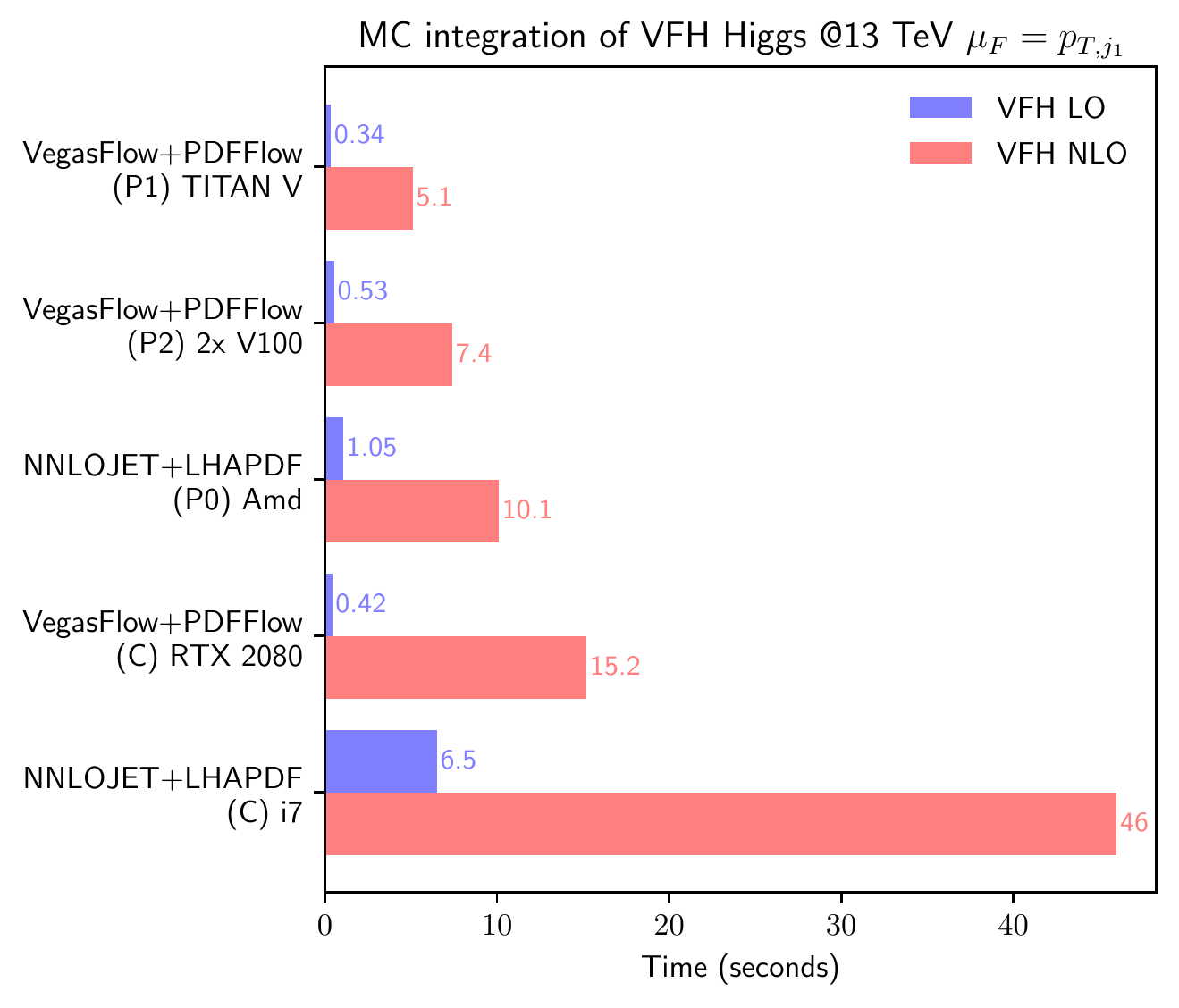}
    \caption{Comparison in time per iteration of
        \vegasflow~\cite{Carrazza:2020rdn, juan_cruz_martinez_2020_3691927} with
        \pdfflow and a Fortran code implementing the same calculation for the
        VFH implementation: NNLOJET.
        we consider the times it take for achieving a per-mille precision with
        both codes at LO and percent at NLO in 10 iterations. The description of
        the systems in which the different codes have been run is given in
        table~\ref{table:hardware}.}
    \label{fig:vfh_lo_nlo}
\end{figure}

Both NNLOJET and our implementation use the Vegas algorithm as integration engine.
We implement the calculation by using VegasFlow~\cite{Carrazza:2020rdn}.

In figure~\ref{fig:vfh_lo_nlo} we show a comparison between NNLOJET and our
\vegasflowpdfflow implementation. We test both codes in consumer-grade (C) and
professional-grade (P*) hardware. We make this distinction as it has been argued
in the past that a standard GPU would not be suitable for NLO calculations due
to physical limitations~\cite{Buckley:2019wov}. We show that this is no longer
the case. A detailed description of the hardware used for this exercise is given
in table~\ref{table:hardware}.

We should note that this is not a true one-to-one comparison as NNLOJET is
heavily optimized for precision and CPU computing which means it includes
strategies to reduce the number of events necessary to achieve target precision.
We anticipate that a full parton-level MC implementation on GPU could achieve
levels of performance far beyond what we obtain in this benchmark.
We also expect a much better efficiency from the point of view of power consumption.

More specifically, as we were able to hold and run a quantity of events of the
order of $10^{6}$ (or bigger) at once in all tested GPUs it is our expectation
that GPU-optimized code will be able to surpass this limit providing even faster
computations even for next-to-next-to leading order integrands or more complex
virtual structures which do not appear in this example.

With all caveats in mind, we obtain a performance equivalent to that of the
CPU-only code when using a consumer grade graphics card (RTX) and surpass it
with professional grade cards such as the Titan V and the V100. In
table~\ref{fig:vfh_lo_nlo} it becomes apparent how the more complicated the
calculation the more there is to gain by using GPUs with increased memory.

\subsection{Multi-PDF members evaluation}

In the context of PDF determination, theoretical predictions for experimental
data points are computed through the convolution of \texttt{FastKernel}
tensors~\cite{Bertone:2016lga} with PDFs evaluated in a grid of $x$ points.

As discussed in~\cite{Carrazza:2019mzf}, we expect performance improvements of
\texttt{FastKernel}-like operations when running parallel multi-PDF member
evaluation on GPU. Such feature is particularly relevant for fitting
methodologies based on the NNPDF methodology, where PDF replicas could be
obtained simultaneously in a single GPU card.

In table~\ref{table:fktable} we show the total evaluation time required to
compute the 13 flavours of NNPDF3.1 NLO for different number of members, $N_{\rm
rep}$. In this example the \texttt{FastKernel} tensors are composed by a total
of 2415 points in $x$, using the P1 system. We observe that \pdfflow on CPU and
GPU times are always smaller when compared to \texttt{LHAPDF} thanks to the
parallel graph evaluation. On the other hand, GPU results are better in the
large $N_{\rm rep}$ regime. We conclude that the multi-PDF member evaluation
implemented in \pdfflow may accelerate computations where a large number of PDF
members and $x$ points are required, thus opening the possibility to perform a
full PDF fit in a single GPU device.

\begin{table}
    \centering
    \begin{tabular}{c|c|c|c}
     $N_{\rm rep}$ &  {\tt LHAPDF} CPU & \pdfflow GPU & \pdfflow CPU\\
     \hline
     10 & 0.08s & 0.07s & 0.05s \\
     50 & 0.41s & 0.35s & 0.28s \\
     100 & 0.83s & 0.69s & 0.56s \\
     200 & 1.87s & 1.46s & 1.12s \\
     300 & 2.85s & 1.29s & 1.79s \\
     400 & 3.63s & 1.69s & 2.12s \\
     \hline
    \end{tabular}
    \caption{Time required to evaluate all 13 flavours from $N_{\rm rep}$
    members of NNPDF3.1 NLO in a grid of 2415 points in $x$, using the P1
    system.}
    \label{table:fktable}
\end{table}

\section{Outlook}
\label{sec:outlook}

Porting PDFs to GPU is an essential step in order to accelerate Monte Carlo
simulation by granting to the HEP community the ability to implement with
simplicity particle physics processes without having to know about the
technicalities or the difficulties of their implementation on multi-threading
systems or the data placement and memory management that GPU and multi-GPUs
computing requires.

\pdfflow is designed to work in synergy with the {\tt LHAPDF} library, therefore
it uses exactly the same PDF data folder structure, and interpolation algorithms
for the PDF and $\alpha_s$ determination. In this work we show that the output
of both libraries is similar even when executing the code on the different
hardware devices.

The current release of \pdfflow has only been tested in GPUs and CPUs, however
we believe that investigation about new hardware accelerators such as Field
Programmable Gate Arrays (FPGA) and Tensor Processing Units (TPUs) could provide
even more impressive results in terms of performance and power consumption.

\section*{Acknowledgements}

We thank Durham University's IPPP for the access to the V100 32 GB GPUs used to
benchmark this code. We also acknowledge the NVIDIA Corporation for the donation
of a Titan V GPU used for this research. S.~C. and J.~C-M. are supported by the
European Research. Council under the European Unions Horizon 2020 research and
innovation Programme (grant agreement number 740006) and by the UNIMI Linea2A
project ``New hardware for HEP''.

\FloatBarrier
\appendix

\section{Comparison between LHAPDF and \pdfflow}
\label{app:pdfacuracy}

\begin{figure*}
    \centering
    \begin{subfigure}{.475\textwidth}
        \includegraphics[width=\textwidth]{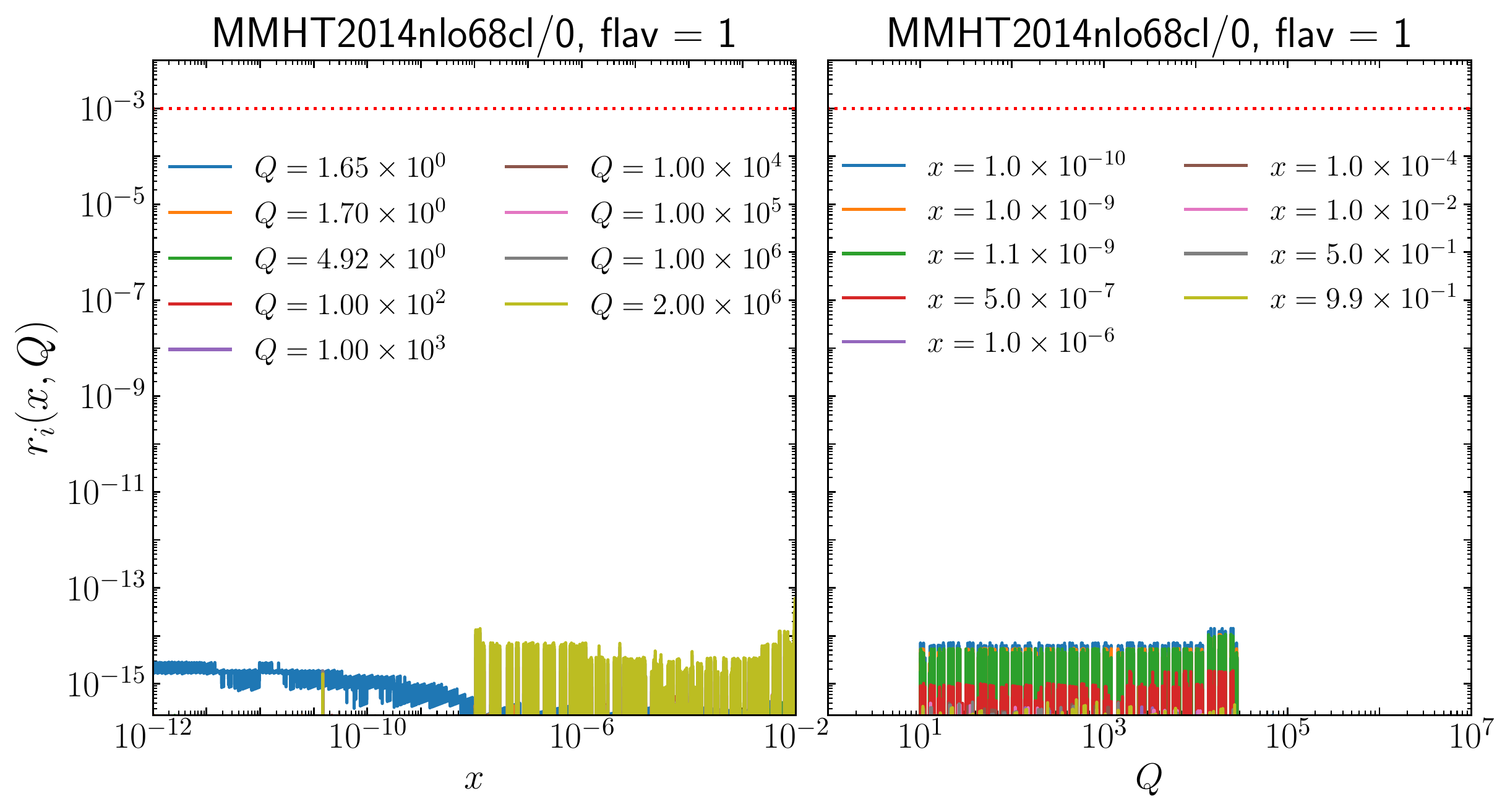}
        \includegraphics[width=\textwidth]{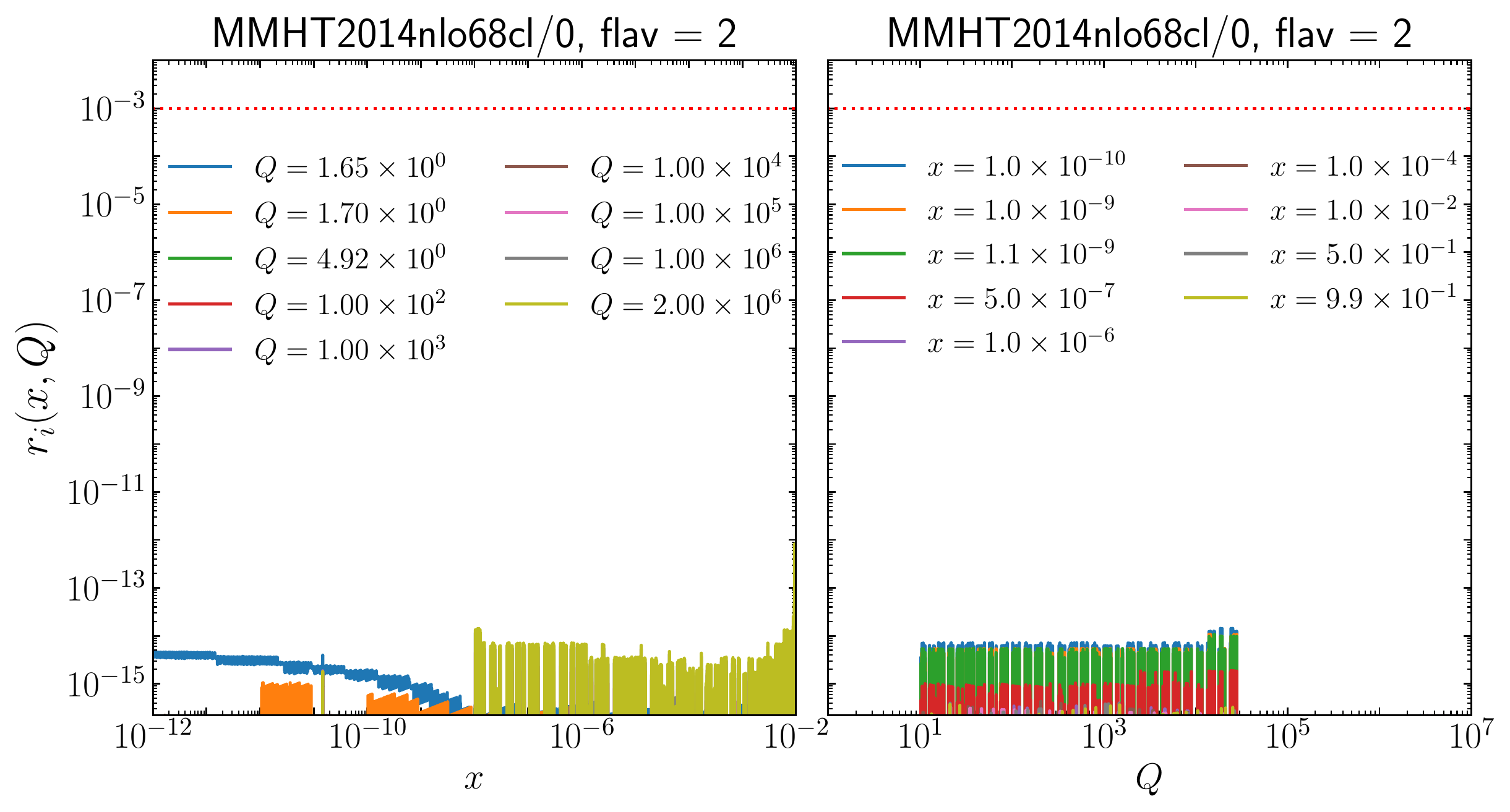}
        \includegraphics[width=\textwidth]{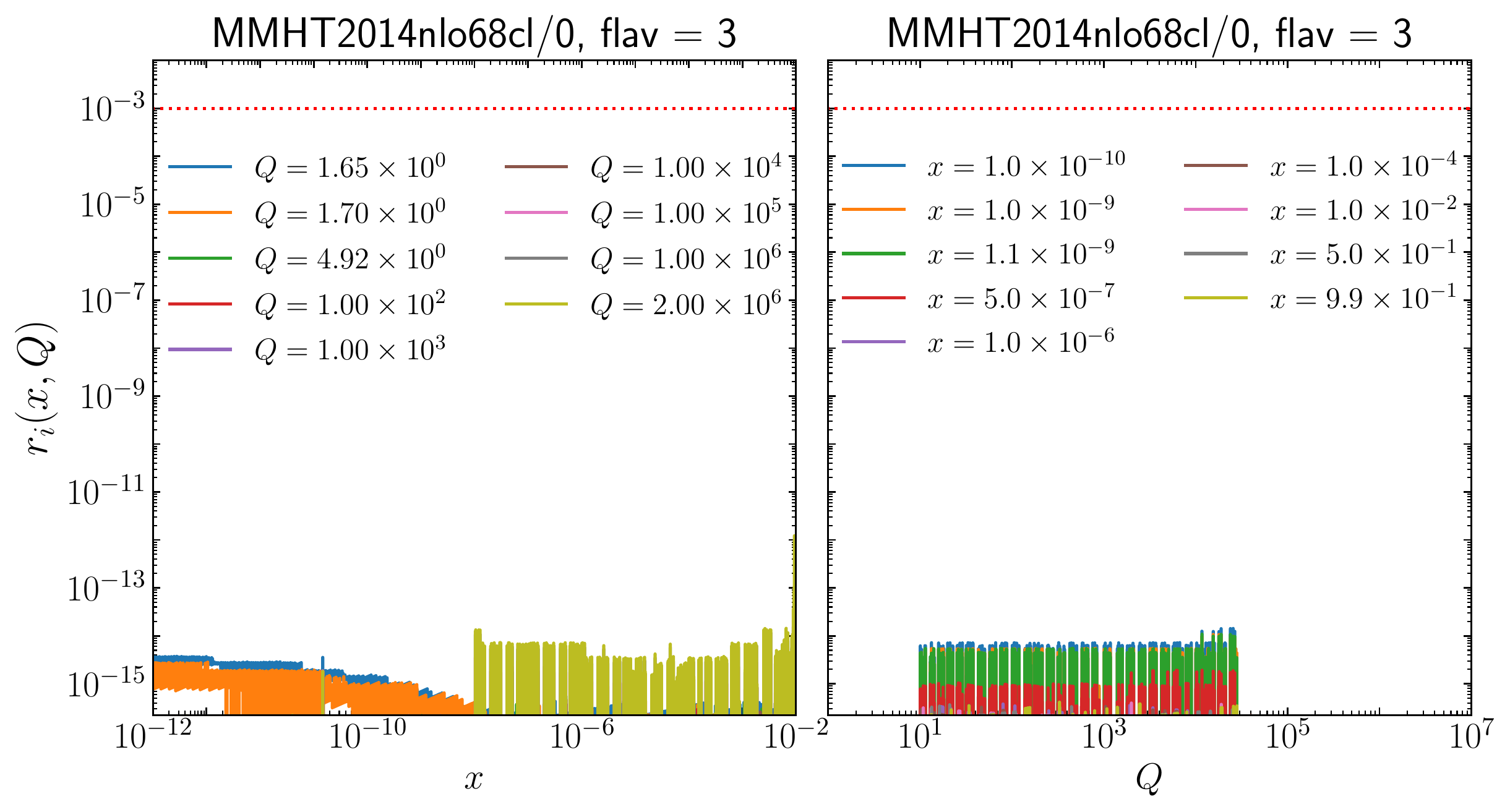}
        \includegraphics[width=\textwidth]{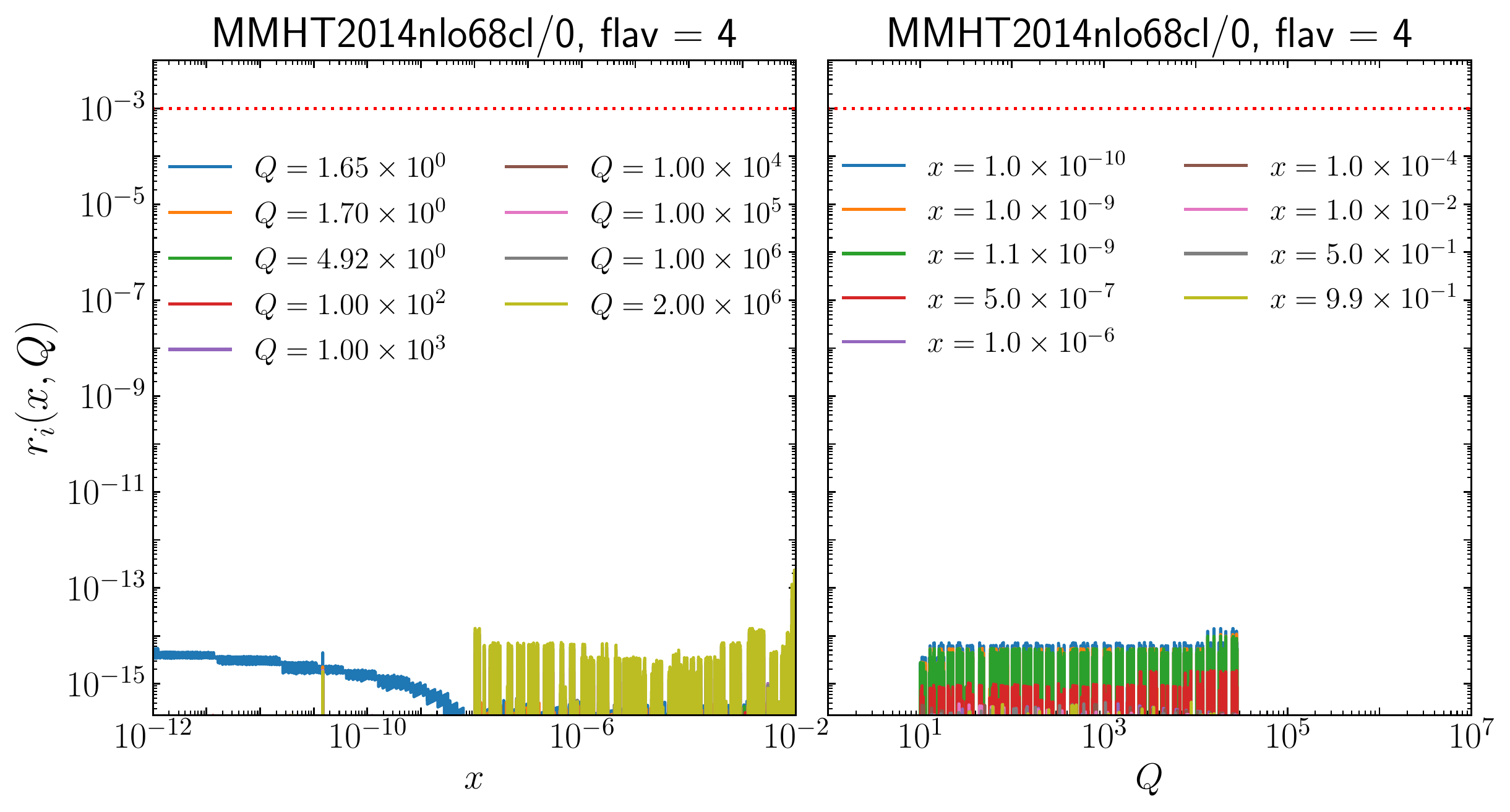}
        \includegraphics[width=\textwidth]{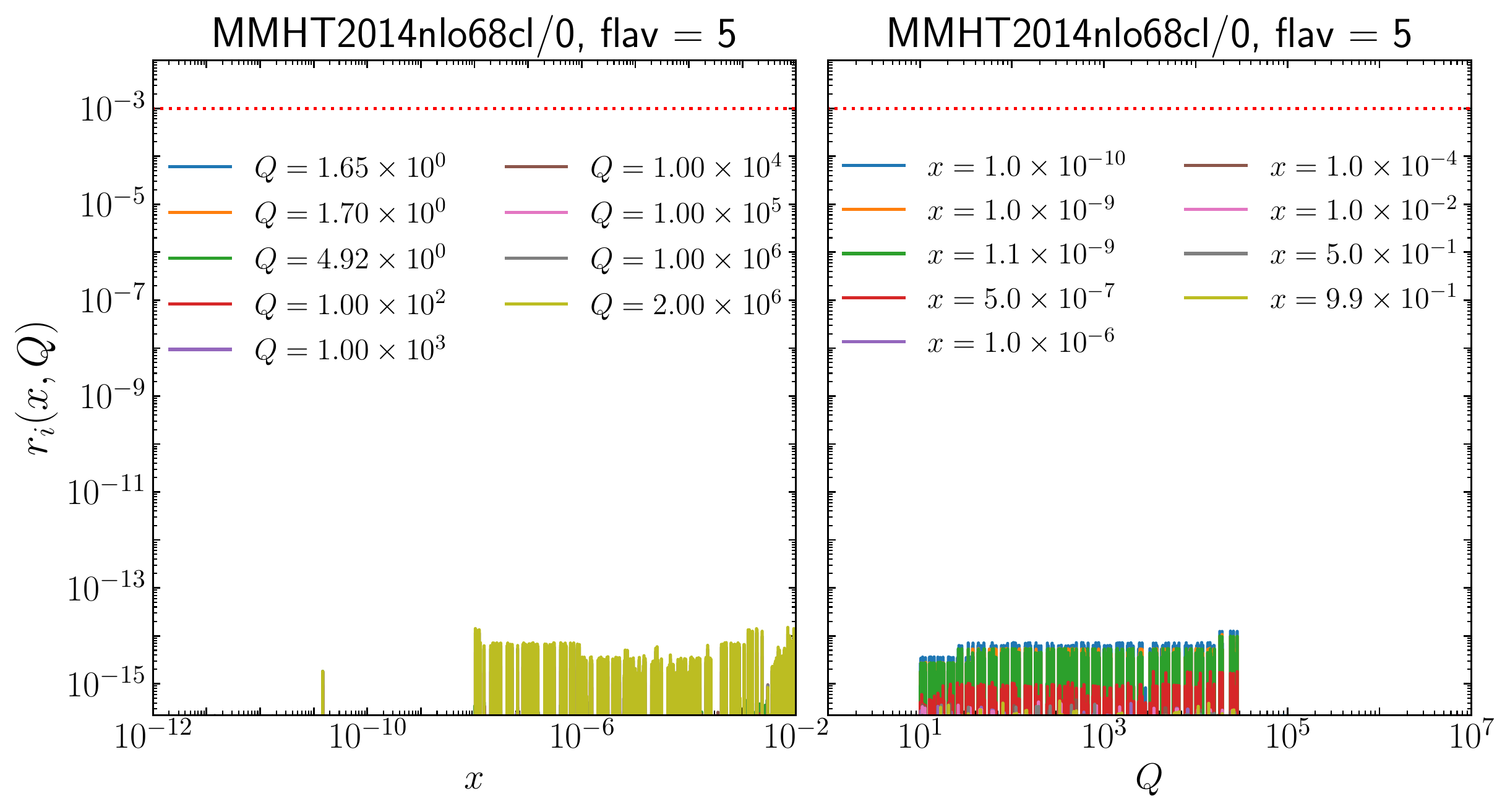}
    \end{subfigure}
    \begin{subfigure}{.475\textwidth}
        \includegraphics[width=\textwidth]{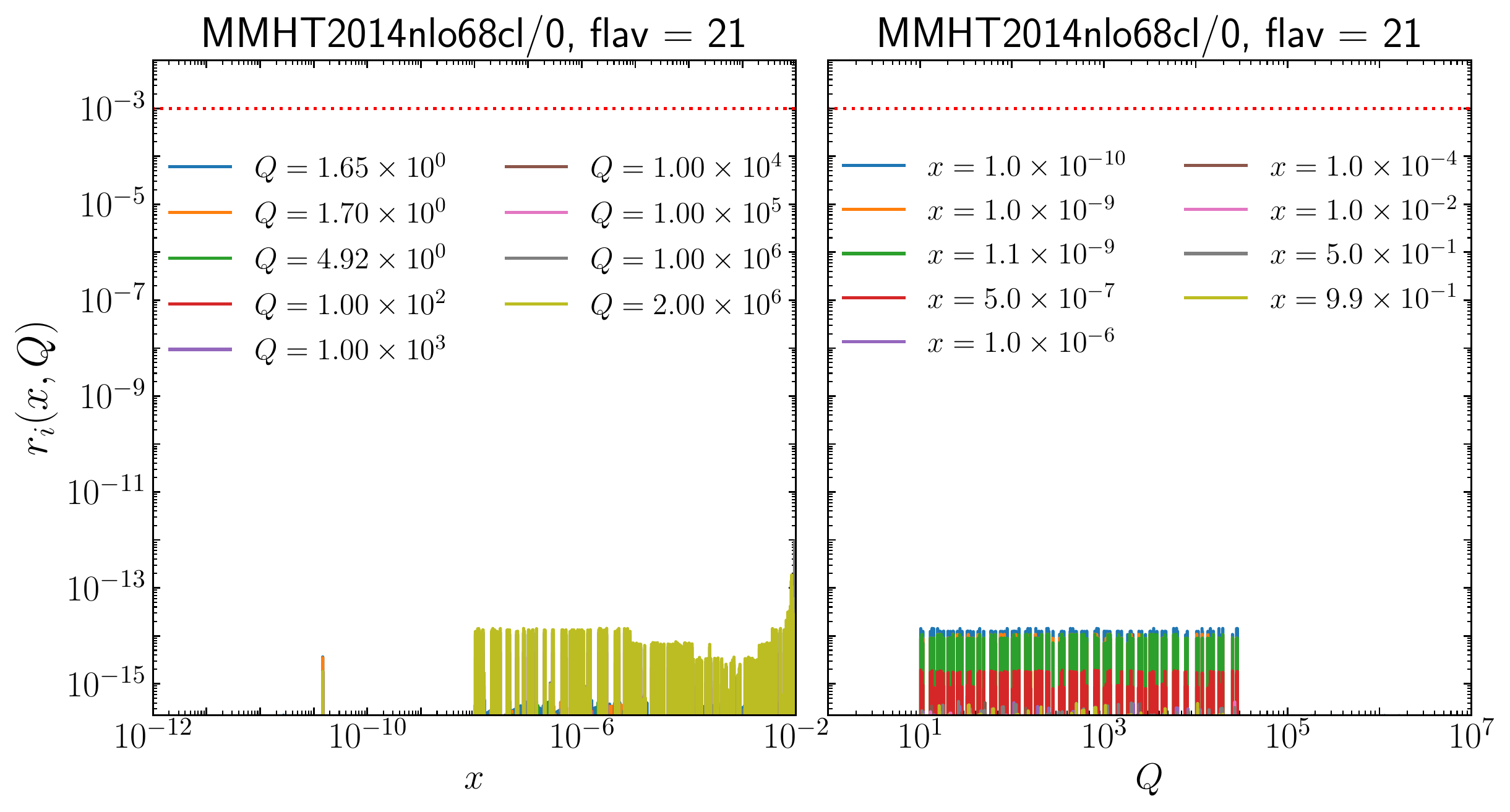}
        \includegraphics[width=\textwidth]{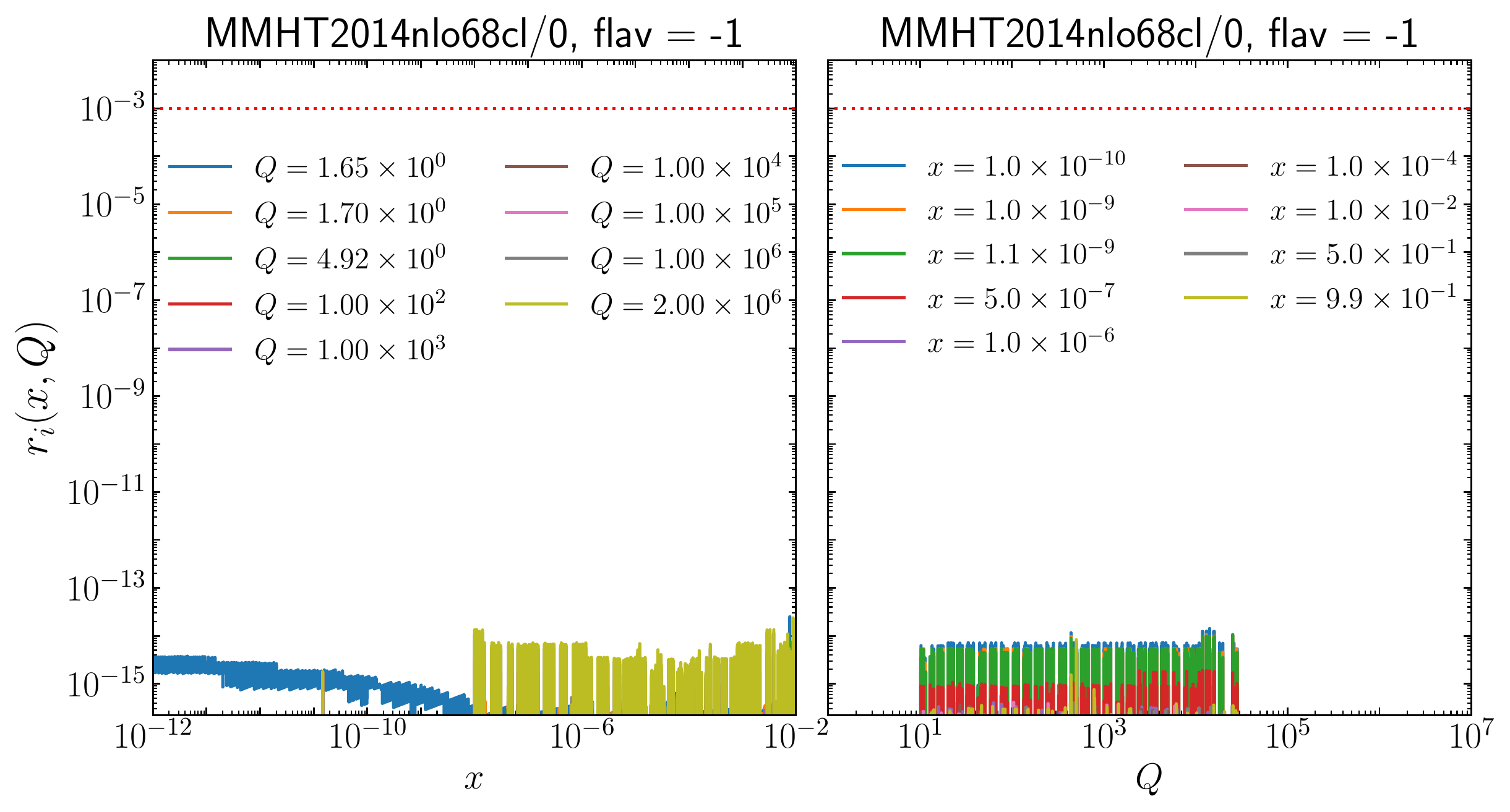}
        \includegraphics[width=\textwidth]{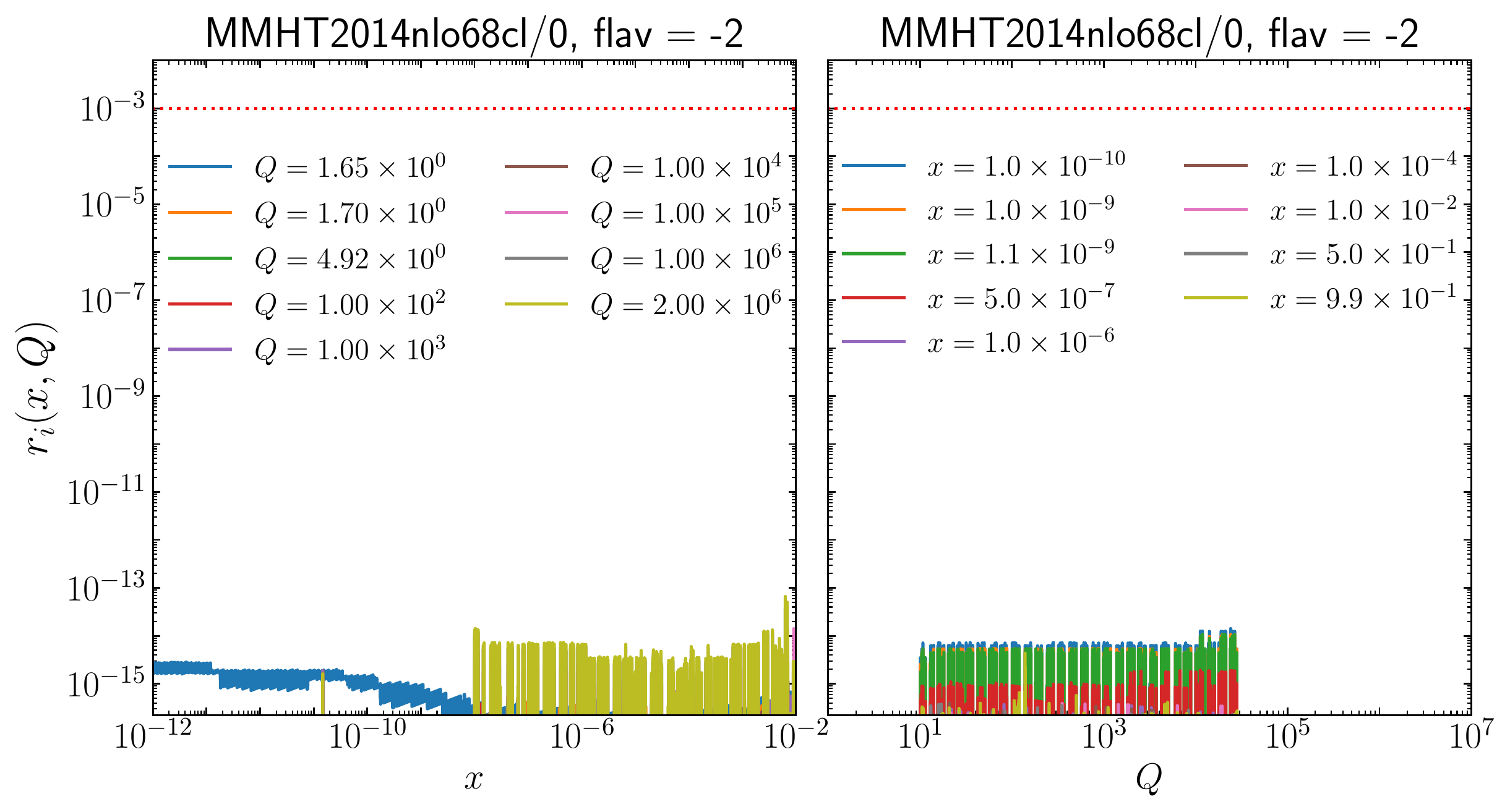}
        \includegraphics[width=\textwidth]{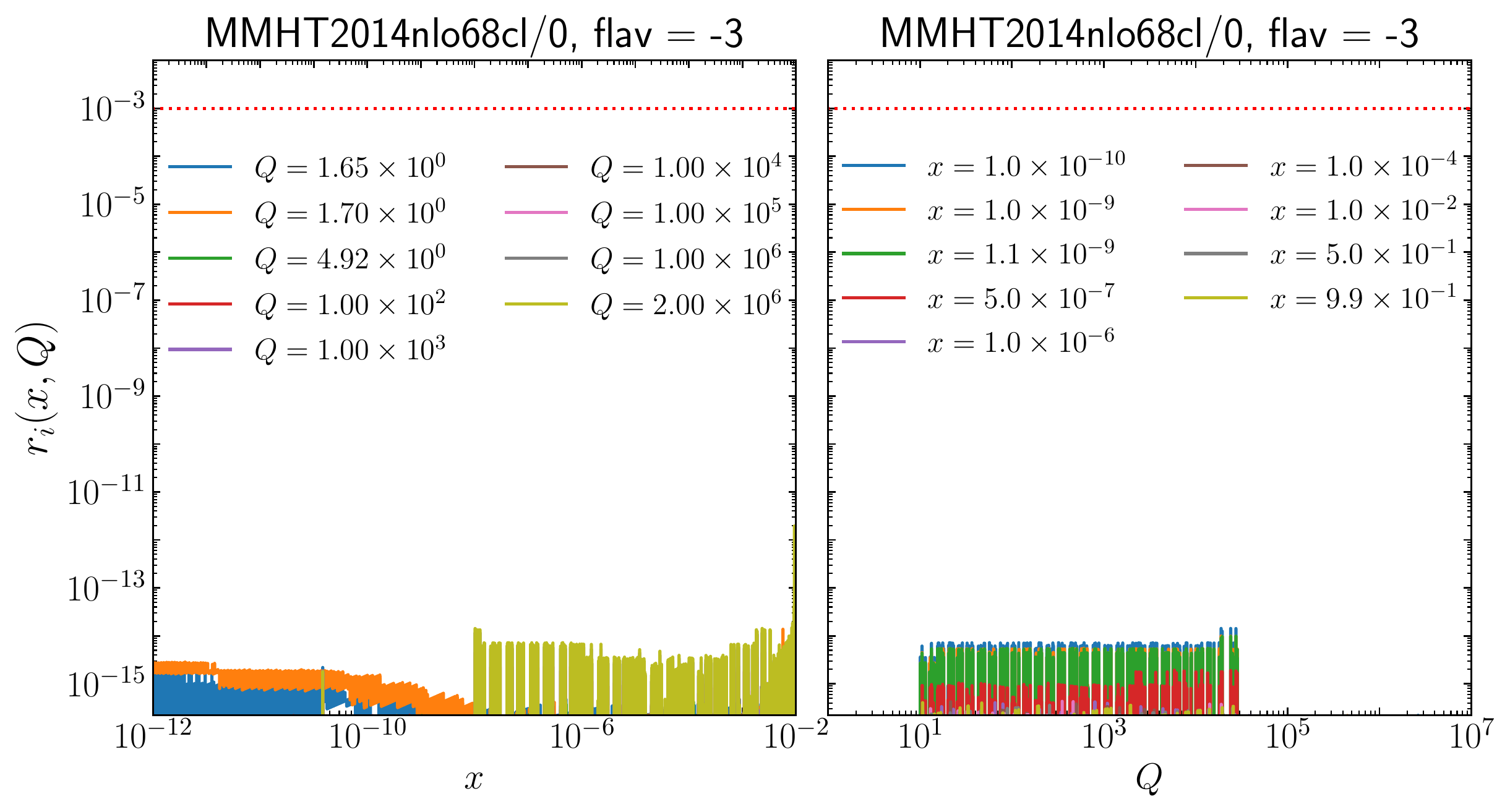}
        \includegraphics[width=\textwidth]{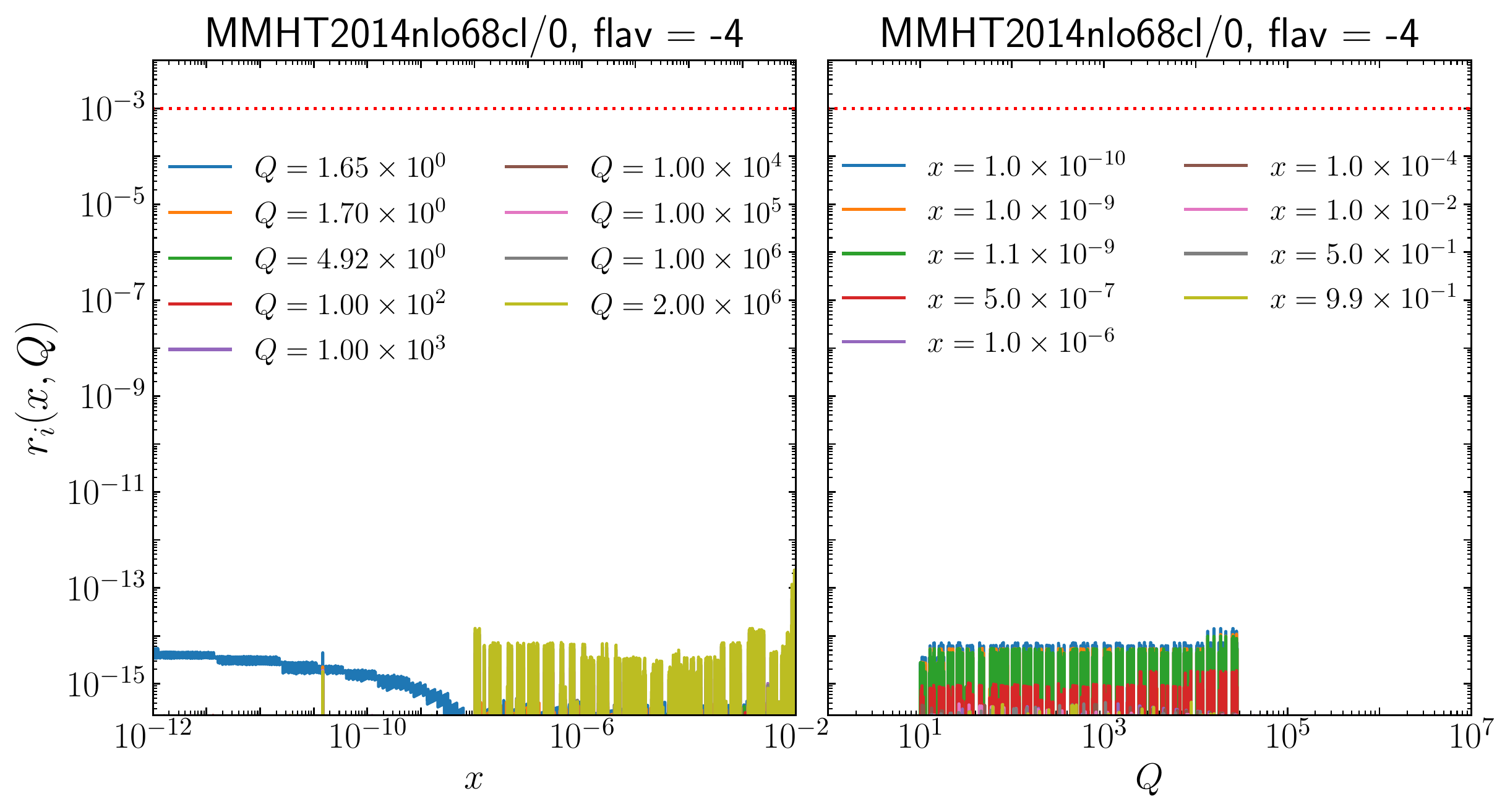}
    \end{subfigure}
    \caption{Relative difference between \pdfflow and LHAPDF (same as~\ref{fig:pdfaccuracy_nnpdf}) for the MMHT2014 NLO
    set for all flavours.}
    \label{fig:pdfaccuracy_mmht}
\end{figure*}

In figure~\ref{fig:pdfaccuracy_mmht} we show the relative difference of
equation~\ref{eq:diff} between \texttt{PDFFlow} and LHAPDF for the MMHT2014 NLO
set for all flavours.

\bibliographystyle{elsarticle-num}
\bibliography{../blbl}

\end{document}